# SUSY Search in Future Collider and Dark Matter Experiments

D. P. Roy

*Homi Bhabha Centre for Science Education, Tata Institute of fundamental Research, Mumbai-400088, India*
*Instituto de Fisica Corpuscular, CSIC–U. de Valencia, Correos E-46071 Valencia, Spain*

**Abstract**. The lightest superparticle in the MSSM is expected to be a Bino, Higgsino or Wino. We consider the dark matter abundance constraint on these LSP scenarios in the minimal SUGRA and AMSB models. We discuss the resulting collider signals for the Bino LSP at LHC and the Higgsino and Wino LSP at CLIC. The Bino, Higgsino and Wino LSP signals in dark matter experiments are also discussed briefly. We conclude with a discussion of these LSP scenarios in nonminimal SUSY models.

## 1. SUSY – PROS AND CONS :

SUSY and in particular the MSSM remains the most popular extension of the standard model because it has four attractive features. It provides (A) a natural solution to the hierarchy problem of EWSB, (B) a natural (radiative) mechanism for the EWSB, (C) a natural candidate for the cold dark matter and (D) unification of gauge couplings at the GUT scale. But it is also known to suffer from two problems. (1) Little hierarchy problem: The LEP limit on the lightest Higgs boson mass requires a large radiative correction from the top - stop loops, which in turn requires a large stop mass, $m_{\tilde{t}} \geq 1 TeV$, i.e. an order of magnitude larger than the EWSB scale. (2) Flavor and CP violation problem: Exptl limit on a FCNC process like µ→γe, getting large SUSY contribution from chargino-sneutrino loops, requires $m_{\tilde{\nu}} \geq 10 TeV$ or degeneracy between the $\tilde{\nu}_e, \tilde{\nu}_\mu$ masses, implying some fine-tuning. Similarly exptl limit on a CP violating process like lectron EDM, getting a large SUSY contribution from neutralino-selectron loop, requires $m_{\tilde{e}} \geq 10 TeV$ or unnaturally small SUSY phases, $\phi_{\mu,A} < 10^{-2}$.

Split SUSY model solves the 2$^{nd}$ problem at the cost of aggravating the 1$^{st}$. In fact the cost is much more, because it assumes the sfermion masses to be many orders of magnitude larger than the TeV scale. This means giving up the solution to the hierarchy problem of EWSB along with the radiative mechanism for the EWSB. This seems to us an unwarrantedly high price to pay, since these were the original motivations for low energy supersymmetry. We shall consider instead a more moderate option of allowing sfermion masses upto the range of several tens of TeV.

# 2. NATURE OF THE LIGHTEST SUPERPARTICLE (LSP) IN THE MSSM :

Astrophysical constraints imply that the LSP must be colourless and chargeless, while the direct dark matter (DM) search experiments disfavour sneutrino LSP. Thus in the minimal supersymmetric standard model the LSP is the lightest neutralino

$$\chi \equiv \chi_1^0 = c_1 \widetilde{B} + c_2 \widetilde{W} + c_3 \widetilde{H}_d + c_4 \widetilde{H}_u \tag{1}$$

The neutralino mass matrix is given by

$$M_N = \begin{pmatrix} M_1 & 0 & -M_Z \sin\theta_W \cos\beta & M_Z \sin\theta_W \sin\beta \\ 0 & M_2 & M_Z \cos\theta_W \cos\beta & -M_Z \cos\theta_W \sin\beta \\ -M_Z \sin\theta_W \cos\beta & M_Z \cos\theta_W \cos\beta & 0 & -\mu \\ M_Z \sin\theta_W \sin\beta & -M_Z \cos\theta_W \sin\beta & -\mu & 0 \end{pmatrix} \tag{2}$$

The diagonal elements are $M_1, M_2, \pm\mu$ in the basis of $\widetilde{B}, \widetilde{W}$ & $\widetilde{H}_{1,2} = \widetilde{H}_d \pm \widetilde{H}_u$, while the nondiagonal elements are less than $M_Z$. The LEP limit on the lightest Higgs mass implies that the diagonal elements are larger than $2M_Z$, at least in the mSUGRA model [1]. Assuming this to hold in more general MSSM would imply small mixings, so that the LSP corresponds approximately to one of these interaction eigenstates - $\widetilde{B}, \widetilde{W} or \widetilde{H}$. There is one exception to this, i.e. when the two lightest diagonal elements are nearly degenerate the mixing angle between them can be large. In this case the LSP can be a mixed $\widetilde{B} - \widetilde{H}$ state. This has been described as the well-tempered neutralino scenario in [2], and a very important example of this is the focus point region of the mSUGRA model [3], discussed below. Apart from such accidental degeneracy, however, the LSP is expected to be a fairly pure state of Bino, Higgsino or Wino.

# 3. DM RELIC DENSITY CONSTRAINTS ON BINO, HIGGSINO & WINO LSP SCENARIOS :

**mSUGRA -** For simplicity let us start with the minimal supergravity (mSUGRA) model, where the SUSY breaking in the hidden sector is communicated to the observable sector by gravitational interaction. Since gravity is colour and flavour blind we have common scalar and gaugino masses and trilinear coupling parameter $m_0, m_{1/2}$ & $A_0$ at the GUT scale. Together with the ratio of the vevs of $H_{u,d} = \tan\beta$ and the sign of the higgsino mass parameter µ we have four and half parameters. The magnitude of µ is fixed by the radiative EWSB condition below. We shall set $A_0 = 0$ and +ve µ, since our results are insensitive to these parameters. Then all the weak scale SUSY masses are given in terms of $m_0, m_{1/2}$ & $\tan\beta$ by the RGE. Since the gaugino masses evolve as their couplings we have

$$M_1 = (\alpha_1/\alpha_G)m_{1/2} \approx 0.4 m_{1/2}, M_2 = (\alpha_2/\alpha_G)m_{1/2} \approx 0.8 m_{1/2}. \quad (3)$$

The weak scale Higgs scalar masses appear in the radiative EWSB condition

$$\mu^2 + M_Z^2/2 = \frac{M_{Hd}^2 - M_{Hu}^2 \tan^2 \beta}{\tan^2 \beta - 1} \approx -M_{Hu}^2 \quad (4)$$

where the last equality holds at tan β > 5, favoured by the abovementioned Higgs mass limit from LEP. The RHS is related to the GUT scale mass parameters by the RGE

$$-M_{Hu}^2 = C_1(\alpha_i, y_t, \tan\beta) m_0^2 + C_2(\alpha_i, y_t, \tan\beta) m_{1/2}^2 \quad (5)$$

where $C_2 \approx 2$ & $C_1 \approx \pm\varepsilon$, thanks to an effective cancellation of the GUT scale value by a negative top Yukawa contribution. For the tan β > 5 region $C_1 = -\varepsilon$, where we have a hyperbolic relation[4]

$$\mu^2 + M_Z^2/2 \approx -\varepsilon m_0^2 + 2 m_{1/2}^2. \quad (6)$$

It is clear from eqs (3) and (6) that over the bulk of the mSUGRA parameter space we have $\mu > M_1$, and hence a bino LSP. But for large $m_0 >> m_{1/2}$, there can be an effective cancellation between the two terms on the RHS of (6) leading to $\mu < M_1$, and hence a higgsino LSP.

Since bino has no gauge charge it can only pair-annihilate via t-channel sfermion exchange, $\widetilde{B}\widetilde{B} \to f\bar{f}$. But the Higgs mass limit from LEP implies $m_{1/2} > 400$ GeV, which in turn implies both $M_1 > 2M_Z$ (as mentioned above) as well as large sfermion masses. The latter makes the pair-annihilation process inefficient, resulting in an overabundance of DM.

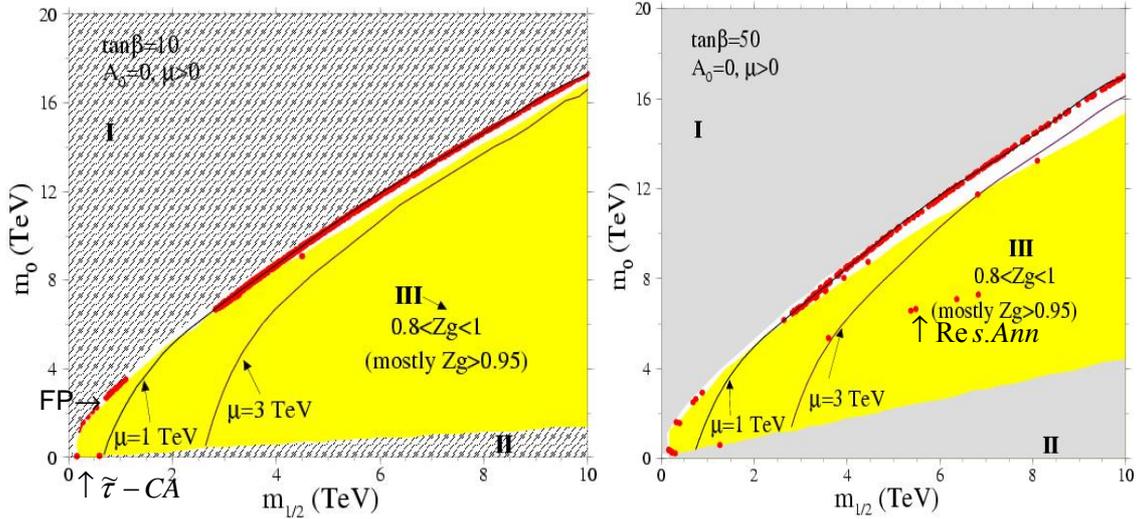

**FIGURE 1.** The mSUGRA parameter space showing the WMAP compatible DM relic density region by red points. The disallowed regions I and II correspond to no EWSB (μ² < 0) and stau LSP regions respectively. The yellow region corresponds to gaugino (essentially bino) dominated LSP [5].

Fig 1 shows that the bulk of the mSUGRA parameter space corresponds to bino dominated LSP (yellow). However most of it is incompatible with the WMAP DM relic density of $\Omega h^2 = 0.09 - 0.13$ [6]. There are only a few compatible points marked red, corresponding to the $\tilde{\tau}$ - CoAnnihilation and the Resonant Annihilation regions. Most of the WMAP compatible points lie in the white strip, corresponding to mixed bino-higgsino LSP (Focus Point region) and dominanantly higgsino LSP ( μ = 1 TeV). $\tilde{\tau} - \text{CoAnnihilation}$ $(\tilde{\tau}\tilde{B} \to \tau\gamma)$ can account for the observed DM relic density at $m_{\tilde{\tau}} \cong M_1$, while Resonant Annihilation $(\tilde{B}\tilde{B} \xrightarrow{A} f\bar{f})$ can account for it at $M_A \cong 2M_1$ & $\tan\beta > 40$. Since the coupling $g_{A\chi\chi} \propto c_{1,2}c_{3,4}$ is suppressed for a $\chi \cong \tilde{B}$ one requires the mass equality for resonance enhancement. The Focus Point region occurs at $\mu \cong M_1$, corresponding to a mixed LSP, $\chi = \tilde{B} - \tilde{H}$. In this case there is efficient pair-annihilation via Z $(\chi\chi \xrightarrow{Z} f\bar{f})$ since $g_{Z\chi\chi} \propto c_3^2 - c_4^2$ is unsuppressed. Note however that in each of these three cases one requires degeneracy between unrelated mass parameters to within 10-15 %, implying some amount of fine-tuning. On the other hand since the charged and neutral higgsinos are degenerate, one has an efficient annihilation channel via W $(\tilde{H}^{\pm}\tilde{H}^0 \xrightarrow{W} WW, f\bar{f})$. Moreover since the higgsino gauge coupling to W is determined by its isospin, the annihilation rate and the resulting relic density are simply determined by the Higgsino mass μ, irrespective of the other SUSY parameters. This accounts for the crowding of the WMAP compatible points on the $\mu \cong 1 TeV$ line.

**mAMSB Model :** It is clear from eq (3) that the wino can not be the LSP if we have unified gaugino masses at the GUT scale. The simplest model for a wino LSP is the minimal anomaly mediated supersymmetry breaking model, where the SUSY breaking in the hidden sector is communicated to the observable sector via the Super-Weyl anomaly term [7]. In fact this contribution is present in all supergravity models; but being a loop contribution it will be small relative to the tree level contribution. So the model assumes absence of the latter. Then the GUT scale gaugino and scalar masses as well as the trilinear couplings are related to the gravitino mass $m_{3/2}$ with coefficients given in terms of the Callan-Symanzik β and γ functions [8], i.e.

$$M_\lambda = \frac{\beta_g}{g} m_{3/2} \Rightarrow M_1 = \frac{33}{5}\frac{g_1^2}{16\pi^2} m_{3/2}, M_2 = \frac{g_2^2}{16\pi^2} m_{3/2}, M_3 = -3\frac{g_3^2}{16\pi^2} m_{3/2}$$
$$A_y = -\frac{\beta_y}{y} m_{3/2} \& m_\phi^2 = -\frac{1}{4}\left(\frac{\partial\gamma}{\partial g}\beta_g + \frac{\partial\gamma}{\partial y}\beta_y\right)m_{3/2}^2 + m_0^2$$
(7)

where g and y represent gauge and Yukawa couplings and $m_0^2$ is a common scalar mass parameter added to prevent slepton mass-squares turning negative when evolved down to the weak scale. Evolving down the gaugino masses to the weak scale including 2-loop contribution gives

$$M_1 : M_2 : |M_3| \approx 2.8 : 1 : 7.1 \tag{8}$$

Like the higgsino the wino LSP can readily pair-annihilate via its isospin coupling to W boson $(\widetilde{W}^{\pm}\widetilde{W}^{0} \xrightarrow{W} WW, f\bar{f})$. In fact since wino is an iso-triplet it has a higher annihilation cross-section than higgsino. Thus one gets the observed DM relic density with a wino mass about twice that of the higgsino LSP, i.e. around 2 TeV.

Fig 2 shows the parameter space of the mAMSB model with the WMAP DM relic density compatible region shown as red dots. The thick column of red dots correspond to a wino LSP of mass $M_2 = 2.1 \pm 0.2$ TeV. One can also see a line of red dots corresponding to a higgsino LSP of mass $\mu \cong 1$ TeV. The physically allowed range of the wino LSP column corresponds to a scalar mass range of $m_\phi = 10 - 30$ TeV.

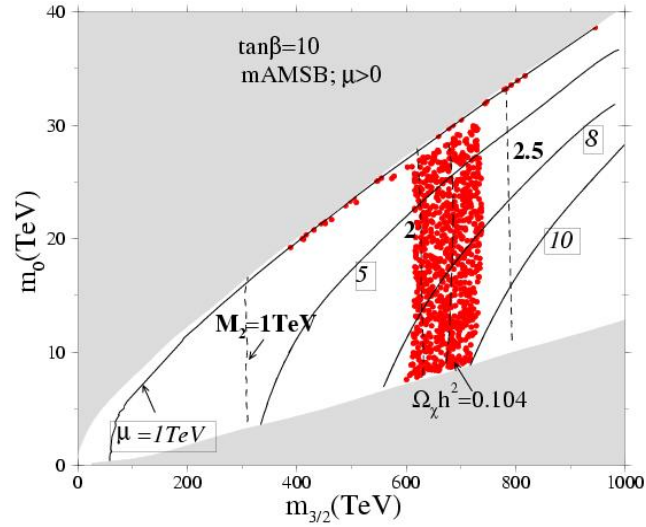

**FIGURE 2.** Parameter space of the mAMSB model with the WMAP DM density compatible region shown as red dots [9]. The upper disallowed region corresponds to no EWSB ($\mu^2 < 0$) while the lower disallowed region corresponds to a slepton LSP (at smaller $m_{3/2}$) or a tachyonic pseudoscalar Higgs boson (at larger $m_{3/2}$).

## 4. BINO LSP SIGNAL AT LHC :

The canonical bino LSP signal at the LHC is the age old multijet + missing-$E_T$ signal coming from [10]

$$\widetilde{q}\overline{\widetilde{q}} \to q\overline{q}\chi\chi \to jj \not{E}_T ; \widetilde{g}\widetilde{g} \to q\overline{q}q\overline{q}\chi\chi \to jjjj \not{E}_T$$

with possibility of additional jets and leptons from the cascade decay of the left-handed (doublet) squarks via wino. This holds throughout the bino LSP region including the Resonant Annihilation region. In the $\widetilde{\tau}$-CoAnnihilation region the wino decays dominantly into $\widetilde{\tau}_1$, followed by the $\widetilde{\tau}_1 \to \widetilde{B}\tau$ decay. Thus one has a τ in addition to multijet + missing-$E_T$. Of course the near degeneracy of the $\widetilde{\tau}$ & $\widetilde{B}$ masses imply the decay τ to be soft. Fortunately it is expected to have polarization $P_\tau \cong +1$, since the $\widetilde{\tau}_1$ is dominantly right-handed. One can exploit this fact to extract this τ signal from the background of negatively polarized τ coming from W decay as well as the fake τ from hadronic jets. In particular in the 1-prong τ-jet channel, requiring the charged prong to carry > 80% of the visible τ-jet momentum effectively suppresses both these backgrounds, while retaining about half the ($P_\tau \cong +1$) τ signal [11]. Using this cut one expects to identify the soft τ signal coming from the $\tilde{\tau}$-CoAnnihilation region.

In the Focus Point region there is a large negative contribution from the top Yukawa coupling to the right-handed stop mass, which automatically leads to an inverted hierarchy of squark masses, as illustrated below [12].

$$m_{\tilde{t}_1}^2 = m_0^2 - \underbrace{y_t}_{2/3} m_0^2 + C m_{1/2}^2 = (1/3) m_0^2 + C m_{1/2}^2 ; m_{\tilde{u},\tilde{d}}^2 = m_0^2 + C m_{1/2}^2$$

$$m_0 = 2 TeV, m_{1/2} = 0.5 TeV \,\&\, \tan\beta = 10$$

$$\Rightarrow m_{\tilde{g}} = 1.3 TeV, m_{\tilde{t}_1} = 1.5 TeV, m_{\tilde{u},\tilde{d}} \geq 2.2 TeV$$

The produced gluino pair is expected to decay dominantly via the relatively light stop, leading to a final state with four b quarks and four W bosons along with the two LSPs, i.e. $\tilde{g} \xrightarrow{\tilde{t}_1} \bar{t} t \chi_i^0, \bar{t} b \chi_j^+ \to 2b 2W \chi ... \Rightarrow \tilde{g}\tilde{g} \to 4b + 4W(\to l) + \not{E}_T ...$

Therefore one expects a distinctive signal containing one or more isolated leptons and four b-jets along with the missing-$E_T$ [12].

## 5. HIGGSINO & WINO LSP SIGNALS AT CLIC :

Unfortunately the higgsino and wino LSP signals can not be seen at the LHC. Their masses are around 1 and 2 TeV respectively, while the corresponding squark and gluino masses are at least 10 TeV (see Figs 1 and 2). Even if the charged higgsino (wino) is pair produced by Drell-Yan process at LHC, the mass degeneracy of the charged and neutral members imply that one sees nothing but a few soft pions in the final state, i.e.

$$\bar{q} q \to \chi^+ \chi^- \to \pi^+ \pi^- \chi \chi ; \Delta m \equiv m_{\chi^\pm} - m_{\chi^0} < 10(0.2) GeV \lrcorner \tilde{H}(\tilde{W})$$

Thus it will be impossible to identify such events at LHC without an effective tag. So one must go to an electron-positron collider with the required CM energy, i.e. the CLIC with the proposed CM energy of 3 to 5 TeV [13]. Here the pair production of higgsino (wino) is tagged by a hard photon from initial state radiation (ISR), i.e.

$$e^+ e^- \xrightarrow{W,B} \chi^+ \chi^- \gamma \qquad (9)$$

Following [14] we shall require the photon to have an angle $\theta_\gamma > 10°$ relative to the beam and to satisfy $E_T^\gamma > E_T^{\gamma\min} \cong \sqrt{s} \sin\theta_{\min}$, which forces one of the outgoing $e^\pm$ from the radiative Bhabha background, $e^+ e^- \to \gamma e^+ e^-$, to come out at an angle $> \theta_{\min}$. Therefore this background can be vetoed by having detectors down to $\theta_{\min}$. This strategy has been followed by the LEP experiments, in particular OPAL [15], to set a mass limit of 90 GeV on Higgsino and Wino. At the CM energy of 3 TeV, $E_T^{\gamma\min} = 50(100) GeV \Rightarrow \theta_{\min} = 1(2)°$.

The OPAL detector has instrumentation down to $2°$; and it seems feasible to extend it down to $1°$ at the future linear colliders [14]. We shall also impose the recoil mass cut

$$M_{rec} = \sqrt{s}(1 - 2E^\gamma/\sqrt{s})^{1/2} > 2m_\chi \tag{10}$$

which is automatically satisfied by the signal (9). The main background is from

$$e^+e^- \to \gamma \nu \bar{\nu} \tag{11}$$

This background was two orders of magnitude larger than the signal (9) at LEP, while it is three (four) orders of magnitude larger at CLIC energies for higgsino (wino). The reason for this background to be so large is that it is dominated by the t-channel W exchange. So it can be suppressed relative to the higgsino signal by having positive(negative)ly polarized electron(positron) beams, which do not couple to W. Fig 3 shows that the resulting background is a factor of 50 larger than the 1 TeV higgsino signal, which is similar to the factor at LEP. Therefore one hopes to extract this signal by looking for the decay pion tracks from $\chi^\pm \to \chi^0 \pi^\pm ...$, as was done at LEP [15].

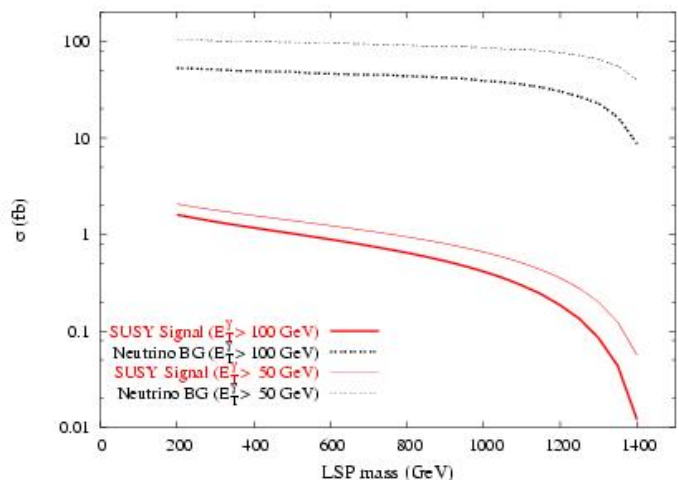

**FIGURE 3.** The higgsino LSP signal (9) and the background (11) cross-sections at CLIC CM energy of 3 TeV with electron and positron beam polarizations of +80% and -60% respectively [5].

The above beam polarizations will suppress the wino LSP signal of eq (9), since it comes only from the s-channel W exchange. On the other hand both the signal and background can be enhanced with opposite beam polarizations. What helps to distinguish this signal from the background is a precise prediction of the charged and neutral wino mass difference, i.e. $\Delta m = 165 - 190 MeV$ [8], which implies $c\tau$ = 3-7 cm. The SLD had microvertex detectors down to 2.5 cm from the beam, while it is expected to go down to 2 cm at the future linear colliders [14]. So one can see the charged wino tracks as two heavily ionising particles along with their decay pion tracks. These will clearly distinguish the signal (9) from the background (11). So the viability of the wino signal is determined primarily by the signal size rather than the signal to background ratio. Fig 3 shows that at the highest CLIC CM energy of 5 TeV and the proposed luminosity of 1000 events/ fb one expects 100 (300) signal events for a 2.1 TeV wino LSP with unpolarized (polarized) beams.

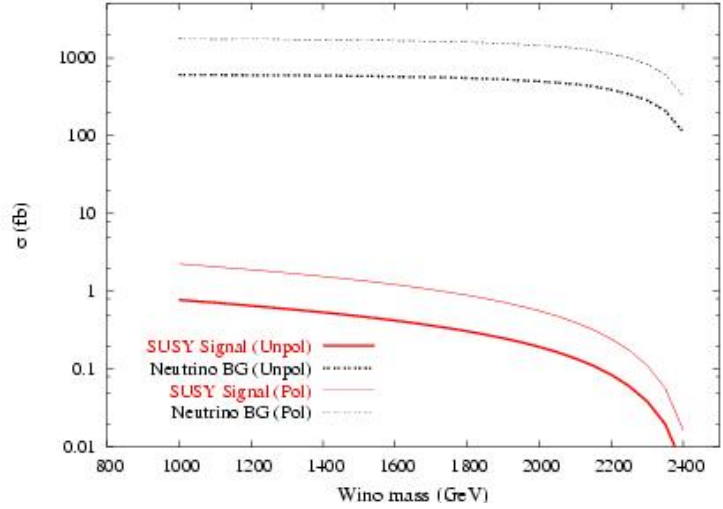

**FIGURE 4.** The wino LSP signal (9) and the background (11) cross-sections at the highest CLIC CM energy of 5 TeV with unpolarized beams as well as with electron and positron beam polarizations of -80% and +60% respectively [9].

It should be noted here that the higgsino (wino) LSP mass less than 1 (2.2) TeV corresponds to an underabundance of DM relic density, which is not actually ruled out by the WMAP data, since there can be other DM candidates as well. Besides there can be non-thermal as well as thermal mechanisms of enhancing the higgsino (wino) DM relic density [8,16]. For this reason we have shown the signal cross-sections in Fig 3 and 4 for lower LSP masses as well.

## 6. BINO, HIGGSINO & WINO LSP SIGNALS IN DARK MATTER DETECTION EXPERIMENTS :

**Direct Detection ( CDMS, ZEPLIN…) –** These are based on the elastic scattering of $\chi$ on a heavy nucleus like Ge or Xe, which is dominated by the spin-independent contribution from Higgs exchange. Since $g_{H\chi\chi} \propto c_{1,2}c_{3,4}$, it is best suited for the mixed bino-higgsino LSP of the Focus Point region. It is less suited for dominantly bino LSP; and totally unsuited for higgsino and wino LSP, which are suppressed by both the $H\chi\chi$ coupling and the large $\chi$ mass.

**Indirect Detection via HE ν from $\chi\chi$ Annihilation in the Sun ( Ice Cube, Antares) –**
At equilibrium the pair annihilation rate is balanced by the trapping rate of $\chi$ in the sun. Since the sun is dominated by H, this is dominated by the spin-dependent contribution from Z exchange, i.e.

$$\sigma_{\chi p} \propto g^2_{Z\chi\chi} \propto (c_3^2 - c_4^2)^2 \qquad (12)$$

This is again well suited for the mixed bino-higgsino LSP of the Focus Point region. It is not suited for the bino or wino LSP, nor for the higgsino LSP, where the two contributions on the right cancel.

**Detection of HE γ Rays from Galactic Centre in ACT ( HESS, MAGIC…..)** – A promising signal of the higgsino (wino) LSP will be the γ rays coming from their pair annihilation at the galactic centre. The largest signal comes from their tree-level annihilation into $WW$, followed by $W \to \pi^0 s \to \gamma s$. Unfortunately it suffers from a large cosmic ray background. A smaller but cleaner signal is provided by the monochromatic γ ray line, coming from the annihilation process, $\chi\chi \to \gamma\gamma, Z\gamma$ via $\chi^\pm W^\mp$ loops. The main source of uncertainty in this signal comes from the assumed profile of DM halo density distribution [17], which can change the result by almost a factor of thousand.

Fig 5 (left) shows the mSUGRA prediction for the line γ ray signal for the generally favoured NFW profile [17] and the typical aperture size ($10^{-3}$ sr) of the atmospheric Cerenkov telescopes [18]. The WMAP compatible 1 TeV higgsino LSP region is clearly identified. Fig 5 (right) shows analogous prediction for the mAMSB model, identifying the WMAP compatible region from a 2 TeV wino and 1 TeV higgsino LSP.

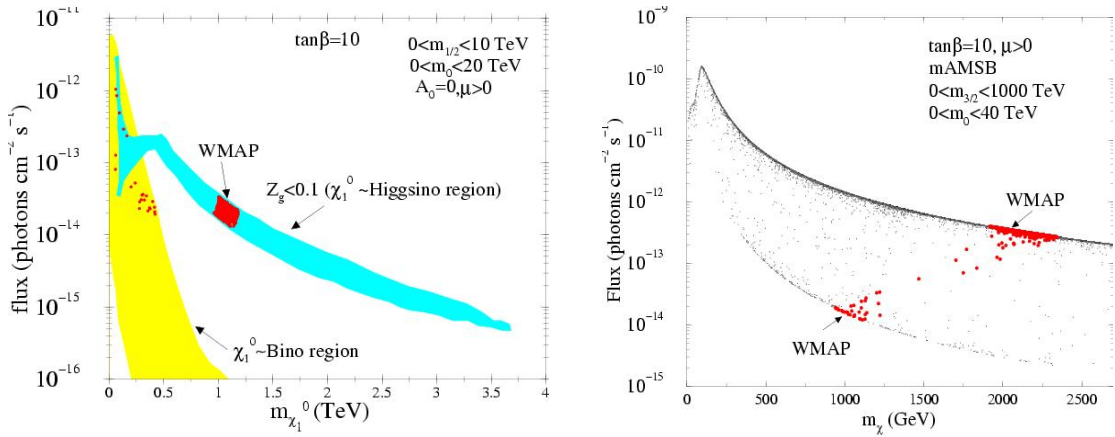

**FIGURE 5.** The predicted line γ ray flux from DM pair annihilation at the galactic centre for the NFW profile in the mSUGRA model (left [5]) and mAMSB model (right [9]). The WMAP DM relic density compatible region, marked red, clearly shows the 1 TeV higgsino and 2 TeV wino LSP regions.

The predicted flux of γ rays is $> 10^{-14}(10^{-13})$ cm$^{-2}$s$^{-1}$ for the higgsino (wino) LSP. In fact the HESS ACT experiment [18] has reported a much higher flux of TeV range γ rays from the galactic centre than this. However the energy spectrum is unlike that of a resolution smeared line γ ray. Instead it shows a power law decrease with energy, which is typical of the other cosmic accelerators like supernova remnants (SNR). Besides that, it is not clear if this signal is coming right from the galactic centre, containing the supermassive black hole Sagittarius A$^*$, or a nearby SNR like Sagittarius A east, lying within the angular resolution of HESS. Therefore it is imperative to improve the energy and angular resolution of the ACT experiments to extract a possible line γ ray signal from this background; and also to look for other possible clumps of DM in the galactic halo.

# 7. HIGGSINO, WINO & BINO LSP IN NONMINIMAL SUSY MODELS :

Finally let us consider the higgsino, wino and bino LSP scenarios in the nonminimal SUSY models. As mentioned above the results of higgsino and wino LSP are independent of the underlying SUSY model, while the bino LSP results are model dependent.

**Higgsino LSP in SUGRA Models with Nonuniversal 1)Scalar & 2)Gaugino Masses** – 1)The higgsino LSP occurs only at the edge of the mSUGRA parameter space because of the small coefficient of $m_0^2$, $C_1 \cong 0$, in eq (5) as the soft breaking Higgs mass-square is effectively cancelled by the negative top Yukawa contribution to the RGE. However in a nonuniversal scalar mass model one can assume this soft breaking Higgs mass-square to be thrice as large as that of the top squark. Since the top Yukawa contribution is proportional to the latter, it implies $C_1 \approx -2 \approx -C_2$ in eq (5), resulting in a small µ and hence a higgsino LSP over a large parameter space [19]. 2) The GUT scale soft gaugino masses come from the vev of the F-component of a chiral superfield S, which is responsible for SUSY breaking, i.e.

$$M_i^G \in \frac{\langle F_S \rangle_{ij}}{M_{Pl}} \lambda_i \lambda_j \quad (13)$$

where $\lambda_{1,2,3}$ are the bino, wino and gluino fields. Since they belong to the adjoint representation of the gauge group the gauge invariance of the above term in the Lagrangian implies that for SU(5) GUT

$$F_S \supset 24 \times 24 = 1 + 24 + 75 + 200 \quad (14)$$

While mSUGRA assumes singlet representation of this superfield, corresponding to universal gaugino masses at the GUT scale, the non-singlet representations would correspond to nonuniversal gaugino mass models. In particular the 200-plet representation implies $M_{1,2,3}^G = (10,2,1) m_{1/2}$. The large value of the bino mass leads to a higgsino LSP over most of the parameter space of this model [20].

**Wino LSP in 1)Nonminimal AMSB & 2)String Models** – 1) The tree level SUSY breaking contributions to the soft scalar masses come from a term analogous to (13), i.e.

$$m_\phi^2 \in \frac{\langle F_S^+ \rangle \langle F_S \rangle}{M_{Pl}^2} \phi^* \phi \quad (15)$$

If one assumes that the SUSY breaking chiral superfield belongs to any representation other than those of (14) then the gauge invariance of (13) and (15) will allow tree level mass for the scalars but not the gauginos. In this case one expects the scalar masses to be typically larger than the anomaly induced gaugino masses by the loop factor of $\sim 10^2$ [21].

2) In String Theory the tree level SUSY breaking masses come only from Dilaton field, while they receive only one-loop contributions from Modulii fields. Thus assuming SUSY breaking by a Modulus field one gets $M_\lambda$ & $m_\phi^2$ at the one-loop level. This leads to $M_2 < M_1 < M_3$ and wino LSP, as in the case of AMSB model, along with $m_\phi \approx 10 M_\lambda$. In fact this model was suggested [22] several years before the AMSB model. Thus in these nonminimal wino LSP models the scalar masses are expected to be in the range of $10^{1-2}$ TeV.

**Bino LSP in Nonuniversal Gaugino Mass Models** – As mentioned above the results of the bino LSP scenario depends strongly on the underlying SUSY model. In particular in the above model of nonuniversal gaugino masses, if one assumes SUSY breaking via a pair of singlet and nonsinglet chiral superfields (belonging to the representations 1+24, 1+75 or 1+200), then it is possible to satisfy the Higgs mass limit from LEP with relatively light sleptons. Consequently one can get WMAP satisfying DM relic density through bino pair-annihilation via the exchange of these light sleptons, i.e. one can recover the so called bulk annihilation region [23]. In this case one also expects a large leptonic branching fraction of SUSY cascade decay via wino at LHC, leading to a robust signature, containing hard isolated leptons along with jets and missing-$E_T$.

## 8. CONCLUSION :

The observed DM relic density has strong constraints on the lightest superparticle of the MSSM, which can be the bino, higgsino or wino. In the case of bino LSP the mass range is constrained to a few hundred GeV, over which one expects very promising signals at the LHC. However, for higgsino and wino LSP the predicted mass ranges go upto 1 and 2 TeV respectively. Moreover it will be hard to detect a higgsino or wino LSP signal at a hadron collider. But they can be detected at an electron-positron collider with the appropriate CM energy, i.e. the proposed mlti-TeV linear collider, CLIC. One also expects a significant flux of monochromatic γ ray lines from their pair annihilation at the galactic centre, at least for the favoured DM halo profiles. They can be detected at future atmospheric Cerenkov telescope experiments, provided the large background of high energy continuum γ rays can be controlled by improved energy and angular resolution. One can also look at other clumps of DM, which are predicted in the galactic halo.

## ACKNOWLEDGMENTS

I gratefully acknowledge financial support from BRNS (DAE) through the Raja Ramanna Fellowship and from MEC through the grants FPA2005-01269, SAB2005-0131.